\documentclass[fleqn,usenatbib]{mnras}

\usepackage{newtxtext,newtxmath}

\usepackage[T1]{fontenc}
\usepackage{comment}

\DeclareRobustCommand{\VAN}[3]{#2}
\let\VANthebibliography\thebibliography
\def\thebibliography{\DeclareRobustCommand{\VAN}[3]{##3}\VANthebibliography}

\usepackage{graphicx}	
\usepackage{amsmath}	
\usepackage{physics}
\usepackage{hyperref}
\usepackage{multirow, tabularx}
\usepackage{array}
\usepackage[table, svgnames, dvipsnames]{xcolor}
\usepackage{makecell, cellspace, caption}
\usepackage{mathtools}
\usepackage{enumitem}
\usepackage{url}
\usepackage{braket}
\usepackage{threeparttable}
\usepackage{orcidlink}
\usepackage{bm}         
\usepackage[normalem]{ulem}

\newcommand\software{\sc}

\defcitealias{Zhang2025UnexpectedClustering}{Z25}

\title[Interpreting UDG clustering by spin bias]{Interpreting the strong clustering of ultra-diffuse galaxies by halo spin bias}

\author[Ma et al.]{
    Qinglin Ma\orcidlink{0009-0000-5148-9457} $^{1}$\thanks{E-mail: maql21@mails.tsinghua.edu.cn},
    Cheng Li\orcidlink{0000-0002-8711-8970} $^{1}$\thanks{E-mail: cli2015@tsinghua.edu.cn}, 
    Yangyao Chen\orcidlink{0000-0002-4597-5798} $^{2,3}$
    and
    Houjun Mo\orcidlink{0000-0001-5356-2419}$^{4}$
\\
$^{1}$Department of Astronomy, Tsinghua University, Beijing 100084, China\\
$^{2}$School of Astronomy and Space Science, Nanjing University, Nanjing, Jiangsu 210093, China\\
$^{3}$Key Laboratory of Modern Astronomy and Astrophysics, Nanjing University, Ministry of Education, Nanjing, Jiangsu 210093, China\\ 
$^{4}$Department of Astronomy, University of Massachusetts, Amherst, MA 01003, USA
}

\date{Accepted XXX. Received YYY; in original form ZZZ}

\pubyear{\the\year{}}

\begin{document}
\label{firstpage}
\pagerange{\pageref{firstpage}--\pageref{lastpage}}
\maketitle

\begin{abstract}
    We use the TNG300-1-Dark simulation to investigate the spin bias of low-mass halos and its connection to the strong clustering of ultra-diffuse galaxies (UDGs) reported by Zhang et al. (2025). By comparing two halo spin definitions—one using only bound particles ($\lambda_{\rm b}$) and another including unbound particles ($\lambda_{\rm a}$)—we demonstrate that the spin bias of 
    low-mass halos critically depends on the definition. While $\lambda_{\rm a}$ yields stronger clustering for higher-spin halos at all masses, $\lambda_{\rm b}$ produces an inverted trend below $M_{\rm h}\sim 10^{11} \rm M_{\odot}/h$. This discrepancy is driven by a subset of halos in high-density environments that have large $\lambda_{\rm a}$ but small $\lambda_{\rm b}$.
    Using an empirical model implemented in SDSS-like mocks, we link the stellar surface-mass-density 
    ($\Sigma_\ast$) of a galaxy to $\lambda_{\rm a}$ of its host halo and find that more diffuse dwarfs 
    tend to reside in higher-spin halos. The model naturally reproduces the observed strong clustering of UDGs within the 
    standard $\Lambda$CDM framework without invoking exotic assumptions such as self-interacting dark matter. 
    The high fraction of unbound particles in UDG hosts likely originates from tidal fields in dense regions, an effect particularly significant for low-mass halos. We discuss how the angular momentum of a halo represented by $\lambda_{\rm a}$ may be transferred 
    to the gas through ram-pressure torquing to affect the size and surface density of the galaxy. 
\end{abstract}




\section{Introduction} \label{sec:introduction}

Dwarf galaxies, defined by stellar masses below $\sim10^9\rm M_\odot$, are the most abundant population of galaxies in the Universe. Their shallow gravitational potentials and low masses make them highly susceptible to both internal feedback processes and environmental effects. Consequently, they serve as crucial laboratories for testing the $\Lambda$CDM paradigm and galaxy formation models \citep{Bullock2017SmallScaleChallenges,Sales2022BaryonicSolutionsChallenges}. A remarkable feature of dwarf galaxies is their structural diversity in size, surface brightness, and morphology; at a fixed stellar mass, their sizes can span more than an order of magnitude \citep{McConnachie2012ObservedPropertiesDwarf}.
Low-surface-brightness galaxies (LSBGs) have attracted significant interest since their discovery in deep, small-area surveys \citep{McGaugh1995MorphologyLSBG,Bothun1997LSBG}, and are subsequently explored in wide-field surveys \citep[e.g.,][]{gellerFaintEndLuminosity2012,venholaFornaxDeepSurvey2017, shieldsJADESLowSurface2025}.
Among these, ultra-diffuse galaxies (UDGs)—characterized by their extremely low central surface brightness ($\mu_{0} > 24$ mag,arcsec$^{-2}$) and large effective radii ($r_{\rm e} > 1.5$ kpc)—have been identified in galaxy groups and clusters \citep[][among others]{vandokkumFortysevenMilkyWaysized2015,Mihos2015GalaxiesExtremes,Koda2015ApproximatelyThousand,Merritt2016DragonflyNearby,Yagi2016CatalogUltra-diffuse,vanderBurg2017AbundanceUltradiffuse,Javanmardi2016DGSATDwarf,Bachmann2021LowSurface,Carleton2023PEARLSStellar,Shen2024DragonflyUltrawide} as well as in the field \citep{Martinez-Delgado2016DiscoveryUltra-diffuse,Roman2017Ultra-diffuseGalaxies,Leisman2017AlmostDarkGalaxies,Zaritsky2019SystematicallyMeasuring,Wei2025ZangetsuCandidate}.
Some UDGs exhibit extraordinary properties \citep{Trujillo2021UltradiffuseCrossroads}. For instance, some of them 
host rich globular cluster systems and have unexpectedly high dynamical masses for their stellar content \citep{vDokkum2016HighStellar,Amorisco2018GlobularComa,Forbes2020GlobularClustersUDGs}, suggesting they may be ``failed'' $L_{\ast}$ galaxies. 
Others appear to be dark-matter deficient based on their velocity dispersion profiles \citep{vDokkum2018GalaxyLacking,vDokkum2019SecondGalaxy}, the global HI profiles \citep{Guo2020} and rotation curves \citep{Sengupta2019DarkHI,ManceraPina2019OffBaryonic,ManceraPina2020RobustHI,Shi2021CuspyDark,ManceraPina2022NoNeedDark,ManceraPina2024ExploringNature}.

Theoretically, the physical origin of UDGs has been extensively studied in recent years, using both semi-analytical models \citep[e.g.,][]{guoDwarfSpheroidalsCD2011,henriquesGalaxyFormationPlanck2015} and hydrodynamical simulations \citep[e.g.,][]{Fitts2017fireField,Jeon2017ConnectingFirst,Revaz2018PushingBackLimits,Grand2021DeterminingFull,Applebaum2021UltrafaintMilky}. UDGs in galaxy clusters are generally attributed to environmental processes, such as tidal stripping and ram pressure in dense environments \citep{Ogiya2018TidalStrippingPossible,jiangFormationUltradiffuseGalaxies2019,liaoUltradiffuseGalaxiesAuriga2019,carletonFormationUltradiffuseGalaxies2019,salasEffectsTurbulenceGalactic2020,tremmelFormationUltradiffuseGalaxies2020}. However, the formation mechanisms of field UDGs remain contentious. Semi-analytical models \citep{Yozin2015QuenchingSurvival,amoriscoUltradiffuseGalaxiesHighspin2016,Rong2017UniverseUltradiffuse,RongUtradiffusesemimodel2024} and some hydrodynamical simulations (e.g., Auriga and TNG50; \citealt{liaoUltradiffuseGalaxiesAuriga2019,benavidesOriginEvolutionUltradiffuse2023}) support the high-spin scenario. This model predicts that UDGs preferentially form in halos with higher spin compared to normal dwarf galaxies of similar stellar mass, aligning with the conventional theory of disk galaxy formation \citep{fallFormationRotationDisc1980,moFormationGalacticDiscs1998}. This scenario is further supported by recent studies of LSBGs in the NIHAO simulation \citep{DiCintio2019NIHAOLow} and the "superthin" galaxies in the IllustrisTNG simulation \citep{Hu2024suerthindisk}, 
with both populations being also found to reside in high-spin halos. 
The apparent discrepancy between feedback-driven and spin-driven origins in the NIHAO simulations reflects a transition in the dominant formation mechanism across different mass scales: while supernova-induced expansion creates low-mass dwarf UDGs ($M_* \lesssim 10^{8.5} \rm{M}_\odot$), high halo spin and angular momentum become the primary drivers for more massive, 'classic' LSB galaxies ($M_* \gtrsim 10^9 \rm{M}_\odot$). 
Such elevated spin parameters could be triggered by major mergers 
that redistribute star formation to the outskirts of galaxies, as seen in the ROMULUS25 \citep{wrightFormationIsolatedUltradiffuse2021,Wright2025MergerDrivenFormation} and the CROCODILE-DWARF \citep{tomaruCROCODILEDWARFAssemblyKinematics2025} simulations, or by an overwhelmingly higher frequency of 
prograde minor mergers that result in the buildup of stellar disk component at radii much beyond those expected  
for normal disk galaxies \citep{Hu2024suerthindisk}.

However, the high-spin scenario has been challenged by other theoretical studies which report comparable halo spin distributions between UDGs and normal dwarfs. For instance, \citet{Zheng2025UltradiffuseGalaxies} found that UDGs in the EAGLE simulation form through high spins in the star-forming gas, which produces extended stellar distributions at large radii, even within halos of normal spin. Alternative formation mechanisms independent of halo spin have thus been proposed. These include supernova feedback-driven outflows that cause expansions in 
both the stellar and dark matter components, as demonstrated in the NIHAO \citep{dicintioNIHAOXIFormation2017,jiangFormationUltradiffuseGalaxies2019}, FIRE-2 \citep{chanOriginUltraDiffuse2018}, Horizon-AGN \citep{martinFormationEvolutionLowsurfacebrightness2019}, and NewHorizon \citep{Jackson2021OriginLSBG} simulations, as well as in empirical models \citep{Freundlich2020ModelCore}. 

More recently, \citealp{Zhang2025UnexpectedClustering} (hereafter \citetalias{Zhang2025UnexpectedClustering}) discovered that isolated, diffuse, blue dwarf galaxies identified from the Sloan Digital Sky Survey \citep[SDSS;][]{2000York} exhibit unexpectedly strong large-scale clustering. This clustering is comparable to that of massive galaxy groups and significantly exceeds predictions based on their halo masses. This finding poses a stringent test for the high-spin scenario and potentially for the standard $\Lambda$CDM model. As demonstrated by the authors, this clustering signature can be reproduced neither by the semi-analytical L-Galaxies model \citep{henriquesGalaxyFormationPlanck2015} nor by the IllustrisTNG hydrodynamic simulation \citep{springelFirstResultsIllustrisTNG2018,pillepichFirstResultsIllustrisTNG2018}, both of which adopt the standard $\Lambda$CDM cosmology. To specifically test the high-spin scenario, the authors applied a subhalo abundance matching (SHAM) model, assuming that, at fixed stellar mass and halo mass, dwarf galaxies in higher-spin halos have larger sizes than their counterparts in lower-spin halos. However, this model fails to reproduce the observed UDG clustering.
Further analysis suggests that the clustering could be explained by the assembly bias inherent in standard $\Lambda$CDM cosmology \citep{gaoAgeDependenceHalo2005}, provided that more diffuse dwarfs form in older, low-mass halos. However, existing galaxy formation models within the standard cosmological framework do not produce this trend. Alternatively, the results can be nicely explained by assuming 
self-interacting dark matter (SIDM; \citealt{Spergel2000ObservationalEvidenceSelfinteracting}; see \citealt{Tulin2018DarkMatter} for a review). In this scenario, self-interactions thermalize the inner halo, leading to cored density profiles and reduced central densities \citep{Ren2019ReconcilingDiversity, Kaplinghat2020DarkMatter}. Since the incidence of self-interaction increases with density and halo age, older halos develop larger cores and lower central densities \citep{jiang2023SemianalyticSIDM}. UDGs residing in older halos would thus exhibit stronger clustering, because older halos are also more strongly clustered than younger halos of the same mass.

The failure of the high-spin scenario to reproduce the clustering of UDGs is attributed to the inversion of the spin bias in the low-mass end. At $z=0$, higher-spin halos with masses above $M_{\rm h} \sim 10^{11.5}\,h^{-1}\mathrm{M}{\odot}$ have a higher bias factor than their lower-spin counterparts of the same mass. However, this trend reverses below this characteristic mass \citep{sato-politoDependenceHaloBias2019,Johnson2019SecondarySpinBias, tucciPhysicalOriginsLowmass2021}. This low-mass reversal was not evident in earlier simulations due to limited mass resolution \citep{bettSpinShapeDark2007,gaoAssemblyBiasClustering2007,faltenbacherAssemblyBiasDynamical2010,salcedoSpatialClusteringDark2018}. In the high-spin scenario, 
UDGs,  which form in low-mass and high-spin halos, would therefore be less clustered than normal dwarfs 
in lower-spin halos due to this bias inversion.
In contrast, other recent high-resolution simulations, such as TNG300-1, do not show such inversion \citep{Montero-Dorta2020Manifestation}. In fact higher-spin halos were found to be more clustered than lower-spin halos of the same mass
across the full mass range probed, from $M_{\mathrm{h}}\sim10^{10.5}~h^{-1}\rm M_{\odot}$ to $M_{\mathrm{h}}\sim10^{14.5}~h^{-1}\rm M_{\odot}$. This absence of inversion was also reported in the ELUCID simulation by \citet{Wang2021EvaluatingOrigins}.
\citet{Montero-Dorta2020Manifestation} noted a key methodological difference: previous studies that found a bias inversion used the {\tt ROCKSTAR} halo finder \citep{behrooziROCKSTARPHASESPACETEMPORAL2012}, whereas TNG300-1 employed a Friends-of-Friends (FoF) algorithm. This suggests that the spin bias of low-mass halos is sensitive to halo definition. The discrepancy in the spin bias 
reported has therefore motivated follow-up studies to investigate its physical origins \citep[e.g.,][]{tucciPhysicalOriginsLowmass2021}.

The primary difference between the {\tt ROCKSTAR} halo finder and the FoF algorithm is that the former utilizes only gravitationally bound particles to define a halo and calculate its spin parameter, whereas the latter uses all particles within the halo's virial radius. Thus, the inclusion of unbound particles in FoF-defined halos seems to be responsible for the difference with {\tt ROCKSTAR}. While using only bound particles is well-motivated for isolated, relaxed systems, this approach may be less suitable for galaxies (e.g. UDGs) that formed in some special environments.   
As suggested by \citet{tucciPhysicalOriginsLowmass2021}, the host halos of UDGs may have experienced tidal interactions with ``splashback halos''—distinct halos that previously passed through the host's vicinity \citep{wangEnvironmentalDependenceCold2007, dalalHaloAssemblyBias2008,hahnTidalEffectsEnvironment2009}. The unbound particles within the virial radius could then originate from stripped matter from these splashback halos during close encounters. Alternatively, 
these particles may be newly accreting material with high specific energy. Both tidal interactions during close encounters and ongoing accretion 
may transfer significant angular momentum to the host halo, thereby altering its spin parameter. 
The strong clustering of UDGs found by \citetalias{Zhang2025UnexpectedClustering} indicates that UDGs are hosted by halos in dense environments 
where such angular-momentum transfer is important. For example, the gas originally attached to 
unbound particles can be decelerated by interaction with the halo gas and become bound to the halo, 
inheriting the specific angular momentum of unbound particles. In such a case, 
including unbound particles in the calculation of the halo angular momentum may be necessary
to estimate the angular momentum of the gas component. 

This study  has two primary objectives. The first is to resolve the reported discrepancies in spin bias by comparing measurements 
that include and exclude unbound particles. The second is to re-evaluate the high-spin scenario for UDGs within the 
$\Lambda$CDM framework by accounting for the dependence of spin on unbound halo particles and the possibility 
of harnessing the angular-momentum associated with unbound particles in dark matter halos. 
In \S\ref{sec:spin_bias}, we show that the conflicting spin bias trends 
primarily stem from different definitions of halo spins. Building on this, 
we then develop an empirical model in \S\ref{sec:empirical-model} using mock catalogs to connect the surface 
mass density of dwarf galaxies with halo spins. Finally, we discuss the broader implications of our 
findings in \S\ref{sec:discussion} and present our conclusions in \S\ref{sec:conclusion}.

\section{Spin bias with and without unbound particles}
\label{sec:spin_bias}

\subsection{IllustrisTNG Simulation}
\label{sec:simulation}

Here, we use the 
TNG300-1-Dark simulation (hereafter TNG300-1-Dark; e.g., \citealt[][]{springelFirstResultsIllustrisTNG2018,pillepichFirstResultsTNG502019,nelsonFirstResultsIllustrisTNG2018,marinacciFirstResultsIllustrisTNG2018,naimanFirstResultsIllustrisTNG2018,Nelson2019IllustrisTNGdata}). The simulation box has a length of 205 $h^{-1}$Mpc on each side, and contains $2500^{3}$ dark-matter particles, each with a mass of $7\times 10^{7} \mathrm{M}_{\odot}$. 
The Planck-15 cosmology \citep{Planck2016CosmologicalParameters}, 
with $\Omega_{\rm m} = 0.3089$, $\Omega_{\rm b} = 0.0486$, $\Omega_{\Lambda} = 0.6911$, $H_0 = 100 h\, {\rm km/s/Mpc}$ with $h = 0.6774$, $\sigma_8 = 0.8159$ and $n_{\rm s} = 0.9667$, is 
adopted. Halos and subhalos are identified by the FoF algorithm  \citep{davisEvolutionLargescaleStructure1985} and \texttt{SUBFIND} \citep{springelPopulatingClusterGalaxies2001}, respectively. Halo merger trees are obtained by the \texttt{SUBLINK} algorithm \citep{Rodriguez-Gomez2015MergerRateGalaxies}.

We use the collisionless dark-matter-only run of TNG300-1 rather than the full hydrodynamic TNG300-1 run. The primary reason is that, as demonstrated in \cite{Zhang2025UnexpectedClustering}, the full hydrodynamic TNG100-1 and TNG50-1 runs fail to reproduce the strong clustering of ultra-diffuse galaxies (UDGs), suggesting that subgrid baryonic prescriptions in these simulations do not yet fully capture UDG formation physics. Our goal is to test the high-spin scenario for UDGs, which requires isolating the role of halo spin from complex baryonic processes. The abundance matching approach adopted in \S\ref{sec:3_2} allows us to do so in a controlled, transparent manner. In addition, we choose the TNG300 volume over TNG100 or TNG500 primarily to minimize the effect of cosmic variance on clustering measurements, while simultaneously ensuring sufficiently high mass resolution.

\subsection{Halo spin parameters}
\label{ssec:spin-properties}

\begin{figure} \centering
    \includegraphics[height=0.45\textwidth]{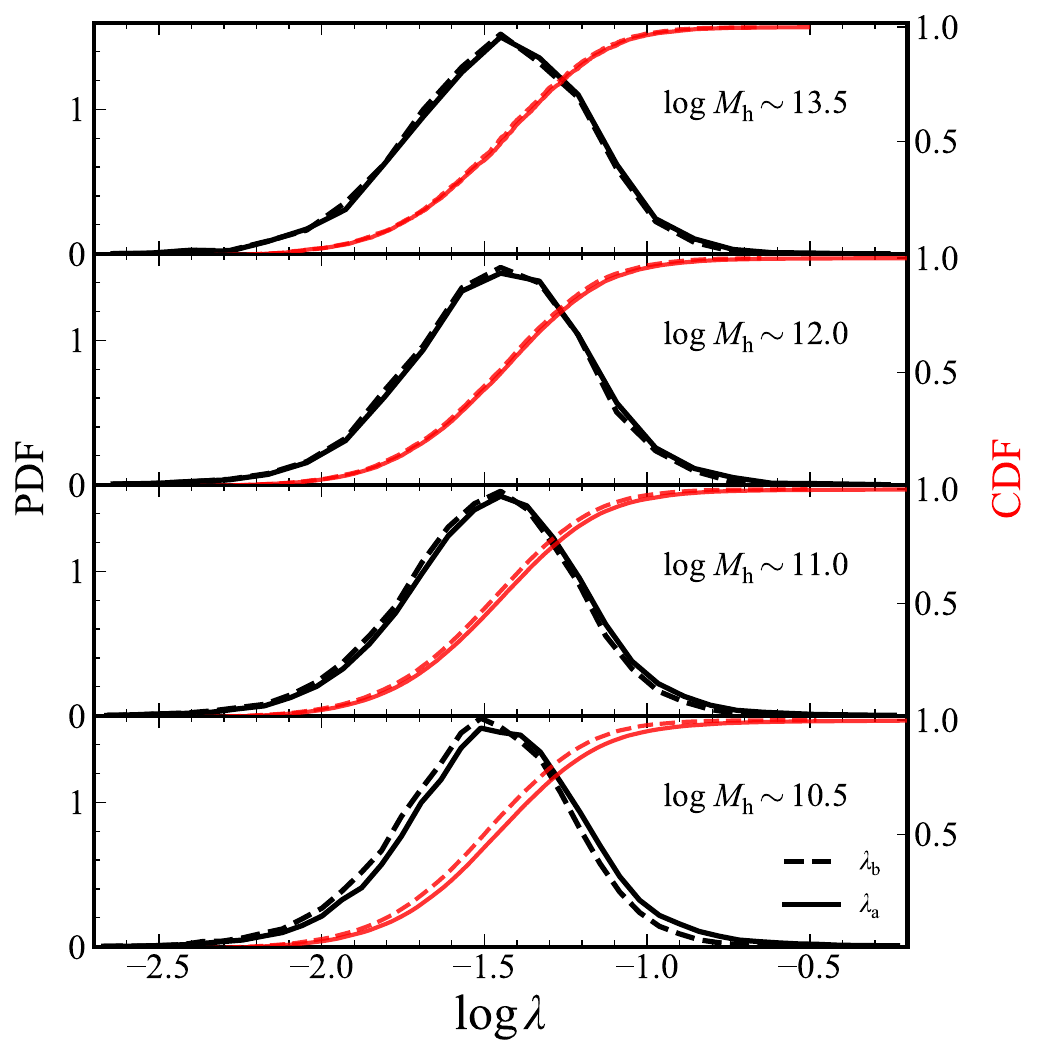}
    \caption{
         \textbf{The probability distribution of halo spin in different mass bins.} In each panel, black lines are probability density functions (PDFs), solid for $\lambda_{\rm a}$ and dashed for $\lambda_{\rm b}$;
         red lines are the corresponding cumulative distribution functions 
         (CDFs), with the $y$-axis on the right. 
    }
    \label{fig:Spin_distribution}
\end{figure}

We define the halo spin parameter $\lambda$ following \citet{bullockProfilesDarkHaloes2001}:
\begin{equation}\label{eqn:lambda}
    \lambda=\frac{\left| \bm{ J }\right|}{\sqrt{2}  M_{\mathrm{h}} R_{\mathrm{h}} V_{\mathrm{h}}},
\end{equation}
where $R_{\mathrm{h}}$ is the virial radius, $\bm{J}$ is the total angular momentum within $R_{\mathrm{h}}$, $V_{\mathrm{h}}$ is the circular velocity at $R_{\mathrm{h}}$, and $M_{\mathrm{h}}$ is the halo mass. 

For each FoF-identified halo in the simulation catalog, we compute two values of $\lambda$: one using only ``bound particles'' ($\lambda_{\text{b}}$) and the other using ``all particles'' ($\lambda_{\text{a}}$) within the virial radius. To identify bound particles, we first calculate $M_{\text{h}}$ and $R_{\text{h}}$ using particles in the vicinity of the halo center (defined as the particle with the 
lowest gravitational potential energy, as provided in the simulation catalog), assuming the virial density threshold \citep{bryanStatisticalPropertiesXRay1998}. For each particle $i$ within $R_{\text{h}}$, we then compute the specific kinematic energy relative to the halo center ($K_i$) and the specific potential energy ($U_i$). In practice, $K_i$ is given by 
\begin{equation}
    K_{i} = \frac{1}{2}\left[\left(\mathbf{v}_{i}-\mathbf{v}_{\mathrm{halo}}\right)+H_{\rm 0}\cdot\left(\mathbf{x}_{i}-\mathbf{x}_{\mathrm{halo}}\right)\right]^2,
\end{equation}
where $H_{\rm 0}$ is the Hubble constant, $\mathbf{x}_{i}$  and $\mathbf{v}_{i}$ are the position and velocity of the $i$-th particle, $\mathbf{x}_{\mathrm{halo}}$ is the position of the halo center, and $\mathbf{v}_{\mathrm{halo}}$ is the halo velocity,  calculated as the average velocity of all particles within $R_{\mathrm{h}}$. To calculate $U_i$, we use all the particles within 4$R_{\text{h}}$ of the halo center: 
\begin{equation}
    U_{i} = -G \sum_{j} \frac{m_{j}}{r_{i,j}},
\end{equation}
where $G$ is the gravitational constant, $m_{j}$ is the mass of the $j$-th particle, and $r_{i,j}$ is the distance between the $i$-th and $j$-th particles. 
We adopt 1 kpc/h as the softening length \citep{springelFirstResultsIllustrisTNG2018}, where $r_{i,j}$ is forced to be 1 kpc/h if it is smaller than this value. 
We set the zero point of the gravitational potential energy at about the turn-around \citep{moGalaxyFormationEvolution2010},
which is typically three to four times $R_{\rm h}$. 
Testing with both $3R_{\rm h}$ and $4R_{\rm h}$, we found the results to be insensitive to the exact 
choice and we chose $4R_{\rm h}$ for our presentation. A particle is considered bound to a halo if $K_i+U_i<0$. 

\begin{figure}
\centering
    \includegraphics[height=0.42\textwidth]{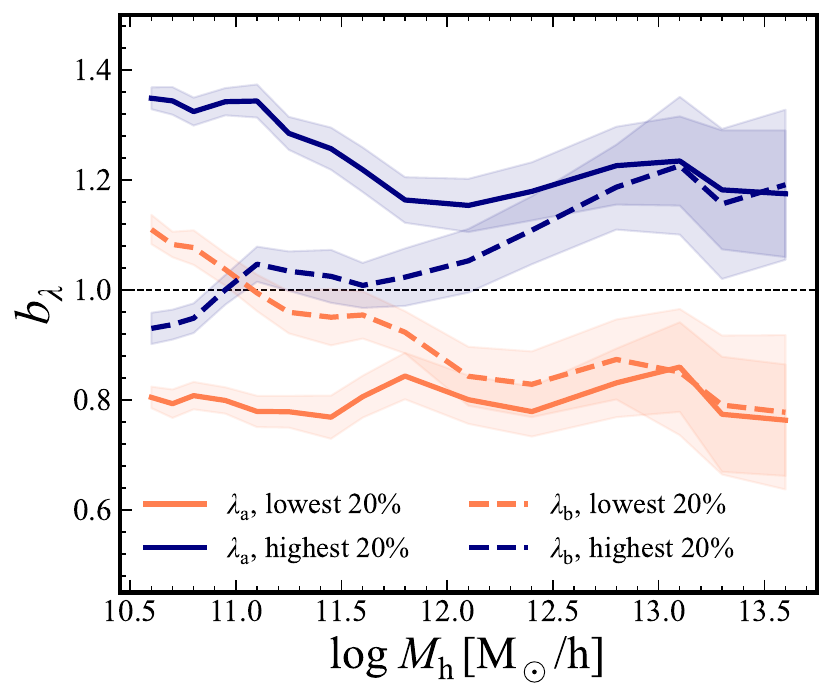}
    \caption{
        \textbf{Spin bias as a function of halo mass.} Dashed lines are for $\lambda_{\rm b}$ and solid lines are for $\lambda_{\rm a}$. The blue and red color corresponds to top 20\% and bottom 20\% in spin, respectively, with the error estimated by 100 bootstrap resamplings. 
    }
    \label{fig:Spin_bias}
\end{figure}

\begin{figure*}
\centering
    \includegraphics[width=0.9\textwidth]{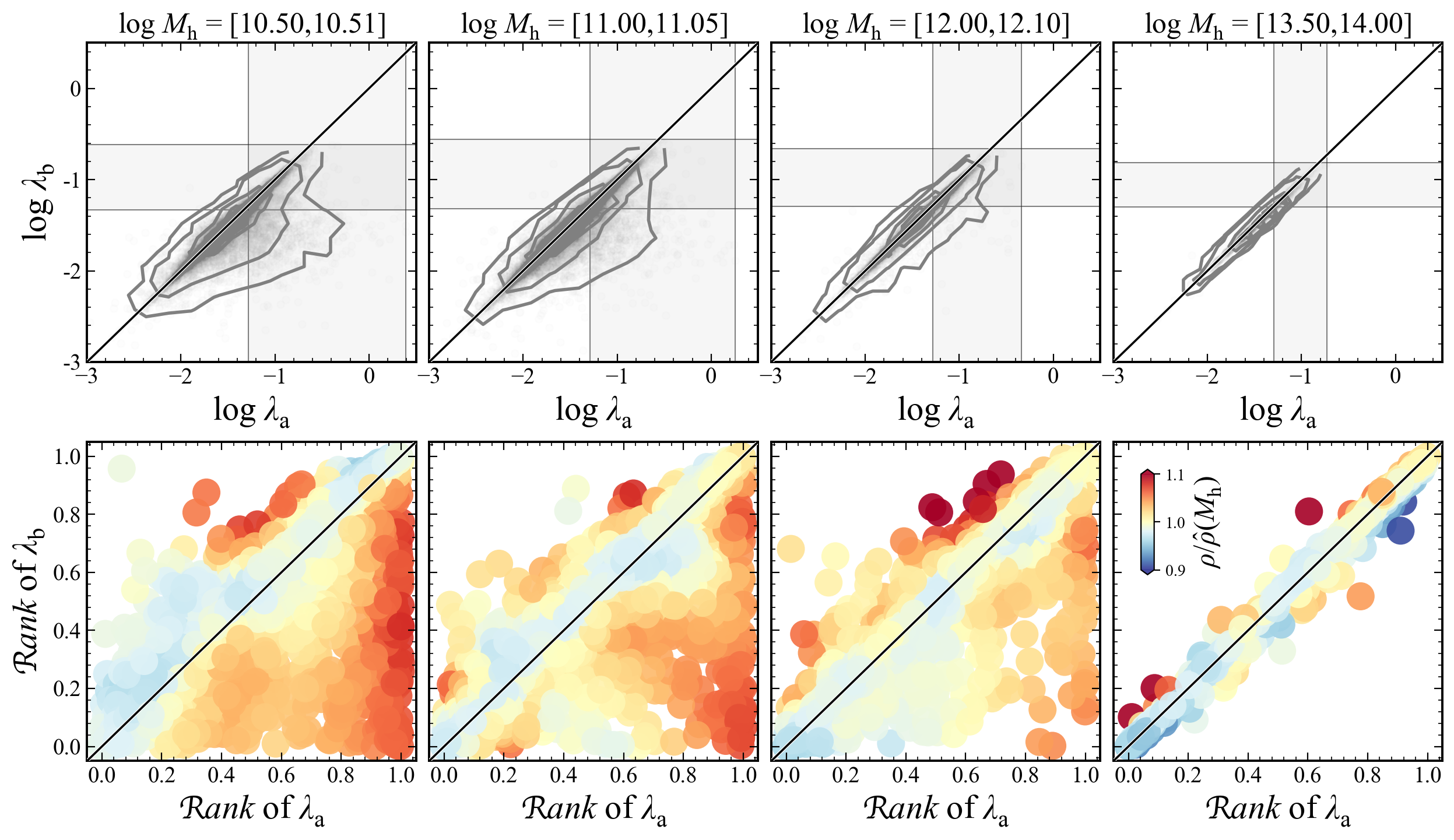}
    \caption{
        \textbf{The distinction of halo spin in different definitions.} 
        \textbf{Top panels:} The spin calculated by all particles ($\lambda_{\rm a}$) versus that calculated by bound particles ($\lambda_{\rm b}$) across different halo mass bins from left to the right panel. 
        The contours cover 75\%, 95\%, and 99\% of the distribution, respectively. 
        The gray shade regions indicate the top $25^{\rm th}$ in $\lambda_{\rm a}$ and $\lambda_{\rm b}$ in each mass bin, respectively. The black line is the one to one line as a reference. 
        \textbf{Bottom panels:} The ranking percentile of halo spin calculated by all particles ($\lambda_{\rm a}$) versus that calculated by bound particles ($\lambda_{\rm b}$) for different halo mass bins from left to the right panel. The color indicates the relative density, defined by the ratio of the number density of this halo with $r \sim 10 h^{-1}$Mpc ($\rho$) to the averaged number density of the whole sample in this mass bin ($\hat{\rho}(M_{\rm h})$), smoothed by the LOESS method \citep{Cleveland01091988}, using the Python package \citep{Cappellari2013}. The black line is the one to one line as a reference.    
    }
    \label{fig:Spin_compare}
\end{figure*}

Using either all particles within $R_\text{h}$ or the subset of bound particles identified above, we recalculate the properties  ($R_{\text{h}}$, $M_{\text{h}}$, $V_{\text{h}}$ and $\bm{J}$) for each halo. We then compute the spin parameters,  $\lambda_{\text{a}}$ or $\lambda_{\text{b}}$, according to \autoref{eqn:lambda}. We select halos with $M_{\rm h} \geq 10^{10.5} h^{-1}\rm M_{\odot}$ to cover the mass range of dwarf galaxies. The lowest-mass halos in our sample contain approximately 800 particles.  
While this number of particles does not guarantee high-precision in the spin measurement\footnote{Theoretically, 800 particles introduce a Poisson error with a variance of $\sigma_{\lambda} \sim 0.01$ \citep{Benson2017ConstrainingNoisefree}, and this noise can dilute the signal.}, our results remain robust against these effects, as demonstrated in subsequent sections.

Fig.~\ref{fig:Spin_distribution} compares the distributions of $\lambda_{\text{a}}$ and $\lambda_{\text{b}}$ in various mass bins. All the distributions follow a log-normal form and exhibit negligible mass dependence, consistent with previous studies \citep{Bailin2005InternalExternalAlignment,Maccio2007ConcentrationEnvironment, bettSpinShapeDark2007,zjupaAngularMomentumProperties2017}. 
Generally, the two spin definitions exhibit very similar distributions, with weak discrepancies emerging only for the lowest mass bin ($\sim 10^{10.5} \, \rm M_{\odot}/h$). In this mass bin, notably, $\lambda_{\rm a}$ presents a more extended high-spin tail while excluding a subset of low-spin halos, and a Kolmogorov-Smirnov (KS) test confirms that the two distributions are statistically distinct with a KS statistic of $0.06$ ($p = 0$) for 70280 low-mass halos. 
These results demonstrate that the discrepancies between spin definitions are generally small and become more pronounced for lower-mass halos. 

\subsection{Spin bias}

For halos of a given mass $M_{\text{h}}$, we estimate the spin bias  as a function of separation $r$ using the following definition: 
\begin{equation}
\label{eqn:spin-bias}
b_{\lambda}\left(r|M_{\mathrm{h}}\right)=\sqrt{\frac{\xi_{\rm hh}\left(r|M_{\mathrm{h}},\lambda\right)}{\xi_{\rm hh}\left(r|M_{\mathrm{h}}\right)}},
\end{equation}
where $\xi_{\rm hh}\left(r|M_{\mathrm{h}}\right)$ is the two-point autocorrelation function of all halos at mass $M_{\text{h}}$, and $\xi_{\rm hh}\left(r|M_{\mathrm{h}},\lambda\right)$ is the two-point autocorrelation function for a subset of halos selected based on their halo spin $\lambda$. Specifically, we rank all halos of a given $M_{\text{h}}$ by $\lambda$, and select the top and bottom 25\% as the high-spin and low-spin subsets. Fig.~\ref{fig:Spin_bias} presents the spin bias for these two subsets as a function of halo mass, derived from the two spin parameters $\lambda_{\text{a}}$ and $\lambda_{\text{b}}$. The values are averaged over the separation range $2 h^{-1}\text{Mpc}<r<10 h^{-1}\text{Mpc}$. 
The uncertainties in $b_\lambda$, indicated by the shaded regions, are given by the variance among 100 bootstrap samples of the whole sample. 

\begin{figure*}
	\centering
	\includegraphics[width=0.95\textwidth]{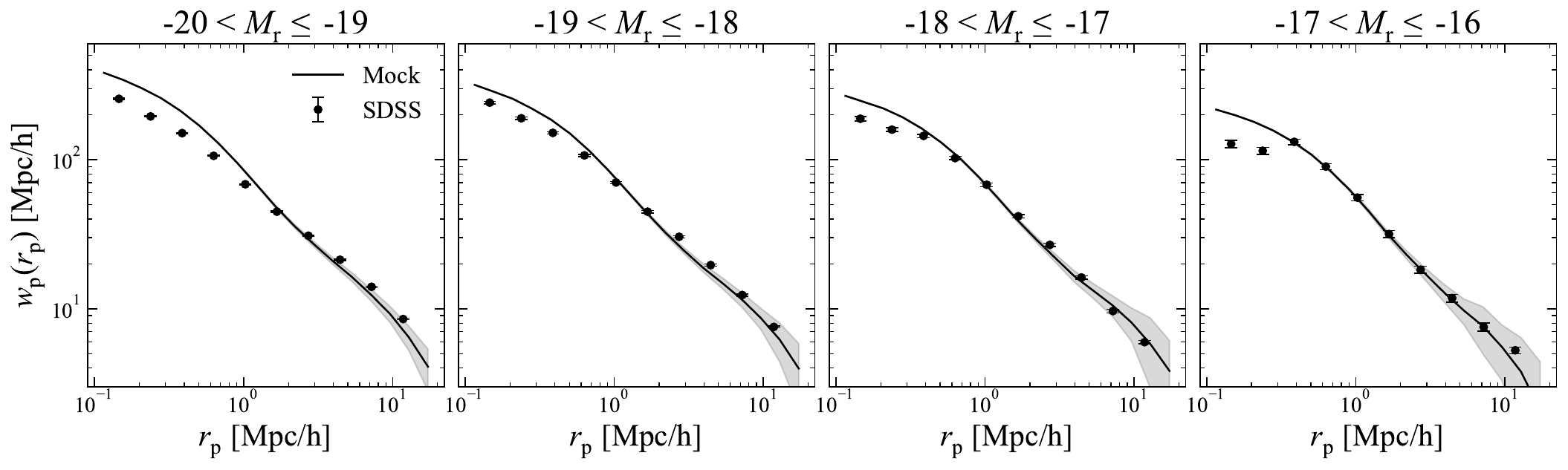}
	\caption{
		\textbf{The projected cross-correlation functions for mock catalogs and SDSS galaxies in different absolute magnitudes from left to the right panels.} 
		For both the SDSS galaxies and our mock catalog, we select the galaxies in given magnitude bin, including the centrals and satellites as the main sample, and utilize the whole sample as the reference sample to calculate the projected cross-correlation function.  
		The error for the SDSS galaxies is estimated by 100 bootstrap resamplings, while error of the mock catalog, showed as the filling lines, represents the 1$\sigma$ scatter between 10 mock catalogs constructed from the TNG300-1-Dark simulation using the same sky mask and magnitude and redshift limits as for the real sample. 
	}
	\label{fig:Mock_2pcf}
\end{figure*}

As shown in the figures, the two spin definitions yield consistent spin biases for high-mass halos ($\log M_{\text{h}}\gtrsim 12.5$). At lower masses, however, the spin parameter derived from bound particles ($\lambda_{\text{b}}$) yields a significantly weaker bias than that obtained from all particles ($\lambda_{\text{a}}$). Furthermore, an inversion of the bias factor in the case of $\lambda_{\text{b}}$ is clearly present at $M_{\text{h}}\sim10^{11}h^{-1}{\rm M_\odot}$. These results are broadly consistent with previous studies that computed spin parameters using {\tt ROCKSTAR} \citep[e.g.,][]{sato-politoDependenceHaloBias2019,Johnson2019SecondarySpinBias, tucciPhysicalOriginsLowmass2021} and the FoF algorithm \citep[e.g.,][]{Montero-Dorta2020Manifestation}, which correspond to our $\lambda_{\text{b}}$ and $\lambda_{\text{a}}$, respectively.

To investigate the discrepancies at low masses, Fig.~\ref{fig:Spin_compare} (upper panels) compares $\lambda_{\rm a}$ and $\lambda_{\rm b}$ in multiple mass bins. The two definitions again produce consistent spin parameters for relatively high-mass halos. At the low-mass end, however, while the majority of halos follow the 1:1 relation, significant deviations are evident for a subset. 
The halos in this subset exhibit large $\lambda_{\rm a}$ while maintaining relatively small $\lambda_{\rm b}$. The most extreme differences approach $\sim$1 dex, indicating that unbound particles contribute substantially to the total angular momentum in these systems, dramatically altering their inferred spin. To further investigate this, we use gray shaded regions to indicate the top 25\% of $\lambda_{\rm a}$ and $\lambda_{\rm b}$ in each mass bin. At low masses, where significant discrepancies in spin bias are seen in the previous figure, a large fraction of halos in the top 25\% of the $\lambda_{\rm a}$ distribution are absent from the top 25\% of the $\lambda_{\rm b}$ distribution, highlighting the definition-dependence in spin ranking.

This effect is more clearly seen in the lower panels of Fig.~\ref{fig:Spin_compare}, which compare the halo rank orders between $\lambda_{\rm a}$ and $\lambda_{\rm b}$. The points are color-coded by the local over-density of halos, estimated from the density within a $10\,h^{-1}\mathrm{Mpc}$ sphere relative to the average density of halos in the same mass bin. Notably, at low masses, halos following the 1:1 relation preferentially inhabit low-density regions. In contrast, halos with the most divergent rankings between the two definitions are predominantly found in higher-density regions and are concentrated in the lower right corner where $\lambda_{\rm a}$ is large but
$\lambda_{\rm b}$ is small. The relative abundance of this population increases systematically with decreasing halo mass, resulting in relatively strong clustering for low-mass halos selected by high $\lambda_{\text{a}}$ or low  $\lambda_{\text{b}}$, as seen in Fig.~\ref{fig:Spin_bias}. Despite their significant effect on the spin bias, these definition-sensitive systems constitute less than 5\% of the total population. Consequently, they produce the high-spin tail in the $\lambda_{\rm a}$ distribution but their impact on the overall spin distributions is limited, as shown previously in Fig.~\ref{fig:Spin_distribution}.

Our analysis demonstrates that the long-standing discrepancy in reported spin bias primarily originates from the different halo spin definitions used in different studies, a point we will discuss further in \S\ref{ssec:origin-spin-bias}. Although different definitions affect only a subset of low-mass halos and have minimal impact on the overall spin distribution, they substantially alter the halo ranking order, thereby inverting the measured spin bias at low masses. The halos experiencing the most dramatic rank changes are predominantly found in high-density environments. This finding underscores the significant role of unbound particles in dense regions, which can substantially enhance the measured halo spin and consequently influence the properties of both the halos and their  galaxies.

\section{Relating the clustering of UDGs to spin bias}
\label{sec:empirical-model}

\begin{figure*} 
	\centering
    \includegraphics[width=0.9\textwidth]{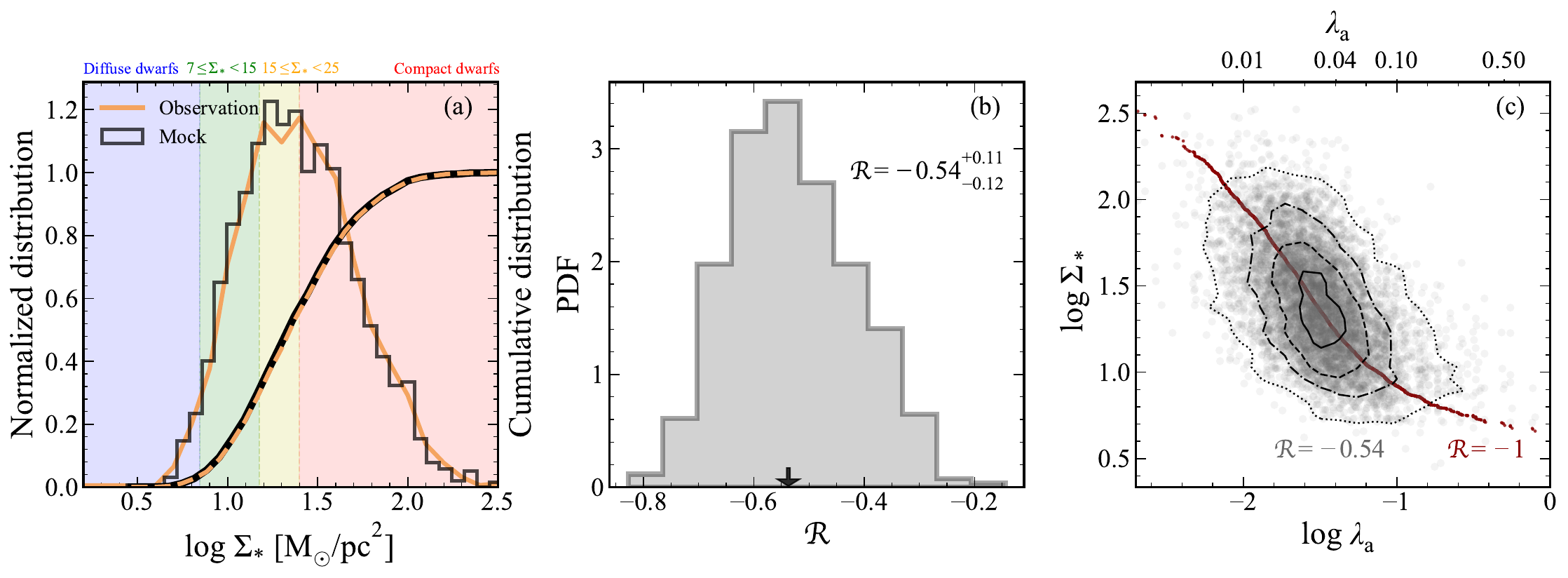}
    \caption{
        \textbf{The main results of the empirical model. }
        \textbf{Panel a:} The normalization distribution (left $y$-axis) and cumulative distribution (right $y$-axis) of the surface mass density $\Sigma_{*}$ of the mock catalog (orange) compared to the SDSS (black). 
        \textbf{Panel b:} the MCMC posterior distribution of the correlation coefficient $\mathcal{R}$ between the surface mass density $\Sigma_{*}$ and the spin $\lambda_{\rm a}$, with the black arrow indicating the median value. The labeled number is the median and $25^{\rm th}$-$75^{\rm th}$ 
        percentile of $\mathcal{R}$.
        \textbf{Panel c:} The $\Sigma_{*}$-$\lambda_{\rm a}$ relation for dwarf galaxies in the best-fit model. The contours enclose 25\%, 50\%, 75\% and 95\% of the entire sample. The red dots correspond to results without adding scatter ($\mathcal{R} = -1$). 
    }
    \label{fig:Mock_dwarf}
\end{figure*}

Motivated by the spin bias inversion at low masses discussed above, we construct an empirical model to populate dark matter halos of different spin parameters with dwarf galaxies of varying surface mass densities. We use the clustering of dwarf galaxies measured from SDSS by \citetalias{Zhang2025UnexpectedClustering} to constrain our model parameters. To ensure a quantitative comparison with observations, we first generate mock catalogs from the simulation that replicate the selection effects of the SDSS galaxy sample. We first describe the construction of these mock catalogs, and then introduce the empirical model for assigning surface densities to dwarf galaxies in the mock catalogs.

\subsection{Construction of mock catalogs}

The mock catalogs are constructed through the following steps. First, under the assumption that each subhalo hosts one galaxy, we assign absolute magnitudes in the r-band ($M_{\rm r}$) to halos using a subhalo abundance matching model  \citep[SHAM;][]{valeLinkingHaloMass2004,guoHowGalaxiesPopulate2010,reddickCONNECTIONGALAXIESDARK2013}. Specifically, we establish a relationship between $M_{\rm r}$ and $V_{\text{peak}}$—the maximum circular velocity attained by a halo during its assembly history, which provides a robust proxy for galaxy luminosity \citep{hearinDarkSideGalaxy2013, reddickCONNECTIONGALAXIESDARK2013}. This relationship is constructed by matching the cumulative number density of halos exceeding a given $V_{\text{peak}}$ threshold to the cumulative number density of galaxies brighter than the corresponding $M_{\rm r}$ threshold. The galaxy number densities are derived from the SDSS DR7 r-band luminosity function (Fig.~9 in \citealt{liConditionalHIMass2022}), incorporating cosmic variance corrections \citep{chenELUCIDVICosmic2019}. To account for both observational uncertainties and intrinsic variations, we introduce 0.5 magnitudes of scatter in the luminosity at fixed $V_{\text{peak}}$. This is implemented by first assigning a provisional magnitude $M^{'}_{\rm r}$ to each halo based on its $V_{\text{peak}}$, then perturbing these values with Gaussian noise ($\sigma$ = 0.5 mag), and finally reassigning $M_{\rm r}$ through rank-order matching of the perturbed magnitudes. Here, the scatter of 0.5 mag is chosen following similar assumptions in the literature \citep[e.g.,][]{trujillo-gomezGalaxiesCDMHalo2011, hearinSHAMBeyondClustering2013}. 

Next, we construct a light cone from simulation snapshots that match the comoving volume and redshift ranges of the SDSS survey, incorporating all relevant observational selection effects. These include redshift-dependent incompleteness (due to the magnitude limit), spatial variations in survey completeness, and K-corrections. \citep{Li2006AGNclustering,mengMeasuringGalaxyAbundance2024}. Following \citetalias{Zhang2025UnexpectedClustering}, we select galaxies with $M_{\rm r} \in [-24,-16]$ and with redshifts $z\in [0,0.2]$ as our complete sample. 
To overcome periodic repetition in stacked simulation boxes, we apply the random tiling scheme \citep{Blaizot2005MoMaFMock}. This allows us to generate multiple independent mock catalogs from a single simulation.
We generate 10 mock catalogs, set by random tiling, with randomized observer positions and lines of sight to estimate cosmic variance. 
We do not consider the effect of fiber collisions in our mock catalogs, as the observational clustering measurements we compare to are revised to account for this effect.
As shown in Fig.~\ref{fig:Mock_2pcf}, the projected cross-correlation functions (PCCFs) with respect to the full galaxy sample, as measured for model galaxies in different intervals of $M_{\rm r}$, show good agreement with the SDSS observations, although the model slightly over-predicts the amplitude of the one-halo terms on small scales. Here, the observational PCCFs of the SDSS/DR7 galaxy sample 
are estimated using the technique described in \citet{liDependenceClusteringGalaxy2006}. 

\subsection{An empirical model to determine surface density of dwarfs}
\label{sec:3_2}


From the mock catalogs, we select galaxies with host halo masses $\log M_{\rm h} [\rm M_{\odot}/h] \in [10.5, 10.8]$ as our dwarf galaxy sample, matching the halo mass range of the observational dwarf galaxy sample from \citetalias{Zhang2025UnexpectedClustering}. In addition,  \citetalias{Zhang2025UnexpectedClustering} excluded dwarf galaxies with red colors and high S\'{e}rsic indices. As suggested by \citet{wetzelGalaxyEvolutionGroups2014}, such galaxies may be hosted by ``splashback'' halos—those that have recently passed through larger host halos and were previously identified as subhalos. We therefore exclude halos identified as splashback halos at any redshift $z < 1$. These constitute approximately 10\% of halos in the mass range $10^{10.5}$--$10^{10.8} h^{-1} \mathrm{M}_{\odot}$, a fraction consistent with that from spherical overdensity (SO)-based merger trees \citep{tucciPhysicalOriginsLowmass2021}. We have repeated our analysis with different upper redshift limits for splashback identification and found that while this choice affects the distribution of our model parameter $\mathcal{R}$ (see below), it does not alter our main conclusions. Our final dwarf galaxy catalog contains between 5988 and 7281 galaxies in each mock sample, with variations due to cosmic variance.

We assume that the dwarf galaxy population follows, on average, a monotonic relationship between galaxy surface mass density ($\Sigma_\ast$) and dark matter halo spin computed using all particles within halo virial radius ($\lambda_{\rm a}$), while allowing for scatter. This relationship is implemented as follows: we first transform the $\lambda_{\rm a}$ distribution into a unit Gaussian distribution, $\mathcal{P}_{\lambda}$. We then introduce a unit Gaussian random variable $\epsilon$ and define a composite variable $\mathcal{P}_{\lambda,\sigma} = \mathcal{R}\mathcal{P}_{\lambda} + \sqrt{1-\mathcal{R}^2}\epsilon$, where the correlation coefficient $\mathcal{R}$ controls the coupling strength. Simultaneously, we map the observed $\Sigma_\ast$ distribution to a unit Gaussian distribution $\mathcal{P}_{R}$. We assign $\Sigma_\ast$ values to dwarf galaxies in the mock catalogs by matching the rank ordering of $\mathcal{P}_{\lambda,\sigma}$ to that of $\mathcal{P}_{R}$. The parameter $\mathcal{R}$ governs the scatter (i.e. Spearman correlation coefficient) in the $\Sigma_\ast$-$\lambda$ relation and can be positive or negative. A negative $\mathcal{R}$ indicates an anti-correlation, where higher spin corresponds to lower surface mass density (i.e., more diffuse galaxies), and vice versa. 
This approach ensures that the resulting surface mass density distribution in our mock catalogs reproduces the observed distribution, as shown in the left panel in Fig.~\ref{fig:Mock_dwarf}.

\begin{figure*}
    \includegraphics[width=0.9\textwidth]{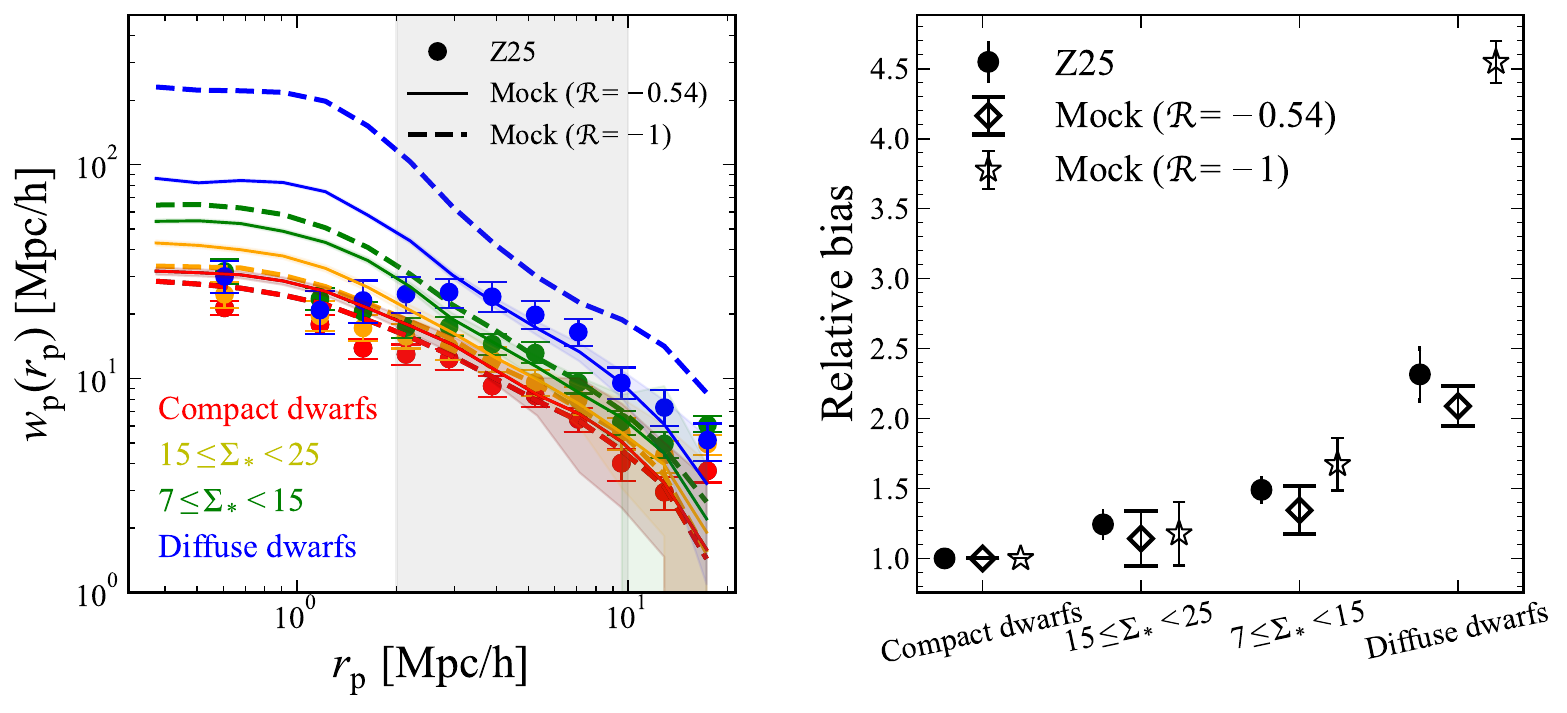}
    \caption{{\it Left:} \textbf{The projected cross-correlation functions of four subsamples for SDSS dwarfs} (dots, \citetalias{Zhang2025UnexpectedClustering}), mock catalog (solid lines) with the best-fitting $\mathcal{R} = -0.54$ and mock catalog (dashed lines) without scatter ($\mathcal{R} = -1$). The error of the mock catalog, only shown in the best-fitting results, is from cosmic variance, estimated by 10 mock catalogs. 
    {\it Right:} Relative bias in four subsamples, with the filled error bar from \citetalias{Zhang2025UnexpectedClustering} for comparison, while the hollow error bar is for the mock catalog with the best-fitting $\mathcal{R} = -0.54$ (diamond) and no scatter $\mathcal{R} = -1.0$ (star).}
    \label{fig:model_clustering}
\end{figure*}

Our empirical model has a single free parameter: the correlation coefficient $\mathcal{R}$ that controls the scatter in the $\Sigma_\ast$-$\lambda$ relation. We constrain this parameter using the observed clustering trends based on surface mass density. Following the partition scheme of \citetalias{Zhang2025UnexpectedClustering}, we divide each of the mock dwarf galaxy samples into four subsamples according to $\Sigma_\ast$: the bottom 2.5\% (most diffuse galaxies), the top 48.3\% (most compact galaxies), and two intermediate bins (48.3\%--74.3\% and 74.3\%--97.5\%). For each subsample, we compute the projected cross-correlation function (PCCF) with respect to the full galaxy sample, $w_{\mathrm{p}}\left(r_{\mathrm{p}}\right)$. Errors are estimated from the standard deviation of $w_{\mathrm{p}}\left(r_{\mathrm{p}}\right)$ among the 10 mock catalogs. We employ Markov Chain Monte Carlo (MCMC) sampling \citep{ForemanMackey2013emceeMCMC} to explore the parameter space and identify the best-fit value of $\mathcal{R}$. The likelihood is defined as $L \propto \exp(-\chi^2/2)$, where
\begin{equation}
\chi^2 = \sum_{i,j} \left[\frac{w_{{\rm p},i}^{\rm obs}(r_{{\rm p},j}) - w_{{\rm p},i}^{\rm mock}(r_{{\rm p},j})}{\sigma_{i,j}}\right]^2.
\end{equation}
Here, $r_{\rm p,j}$ denotes pair separations in the range 2–10 $h^{-1}\rm Mpc$, $w_{{\rm p},i}^{\rm obs}$ and $w_{{\rm p},i}^{\rm mock}$ are the PCCFs of the $i$-th subsample from observations \citepalias{Zhang2025UnexpectedClustering} and our mock catalogs (using the mean across 10 mocks), and $\sigma_{i,j}$ is the combined error from the observational uncertainties and the standard deviation across our mock catalogs. The posterior distribution of $\mathcal{R}$ from our MCMC fitting is presented in the middle panel in Fig.~\ref{fig:Mock_dwarf}. The median value of $\mathcal{R} \approx -0.54$ indicates a negative correlation between surface mass density and halo spin, suggesting that more diffuse dwarf galaxies form in higher-spin halos. The right panel in Fig.~\ref{fig:Mock_dwarf} displays the $\Sigma_\ast$-$\lambda_{\text{a}}$ relation for the dwarf galaxies in the best-fit model with $\mathcal{R}=-0.54$. The anti-correlation between the two parameters and the relatively large scatter are clearly seen from the figure. 

The best-fit PCCFs for the four subsamples are shown in the left panel of Fig.~\ref{fig:model_clustering}. The relative bias, shown in the right panel, is defined as the ratio between the PCCF of each subsample and that of the subsample of the most compact dwarfs, computed over the projected separation range of 2--10 $h^{-1}$Mpc. Observational measurements from \citetalias{Zhang2025UnexpectedClustering} are plotted in both panels for comparison. As expected, the large-scale clustering and the relative bias are both well reproduced by the model. It is noticeable that, on scales below $\sim$1 $h^{-1}$Mpc, the observed  $w_{\mathrm{p}}\left(r_{\mathrm{p}}\right)$ exhibits a clustering amplitude 
that is comparable across different subsamples. In contrast, the predicted correlation function on small scales 
shows a significant dependence on $\Sigma_\ast$, similar to that at larger scales. As demonstrated by \citetalias{Zhang2025UnexpectedClustering}, the measurements of the small-scale clustering is particularly susceptible to effects of satellite contamination in the observational sample.
We therefore focus our analysis on large scales, specifically over scales between 2-10 $h^{-1}$Mpc.

One might question whether the scatter in the $\Sigma_\ast$-$\lambda_{\text{a}}$ relation is necessary. A relation with zero scatter corresponds to our model with $\mathcal{R}=-1$, which is plotted as the red line in the right panel in Fig.~\ref{fig:Mock_dwarf}. However, as shown in the middle panel in Fig.~\ref{fig:Mock_dwarf}, the posterior distribution of $\mathcal{R}$ indicates that the probability for $\mathcal{R}\lesssim -0.8$ approaches zero. The PCCFs and relative bias factors for the $\mathcal{R}=-1$ model are plotted as dashed lines and stars in Fig.~\ref{fig:model_clustering}. This model significantly overpredicts the large-scale clustering and relative bias for the most diffuse galaxy subsample. While the $\mathcal{R}=-1$ model shows reasonable agreement with observations for intermediate and low values of $\Sigma_\ast$,
it likely reflects the weak dependence of clustering on $\Sigma_\ast$ in these bins rather than validating a zero-scatter relation. The scatter in the $\Sigma_\ast$-$\lambda_{\text{a}}$ relation may indeed vary with $\Sigma_\ast$, but current clustering measurements alone cannot constrain such variations. For this reason, we have assumed a constant $\mathcal{R}$ over the full range of $\Sigma_\ast$.

\section{Discussion}
\label{sec:discussion}

\subsection{Understanding the discrepancy in spin bias at low masses} 
\label{ssec:origin-spin-bias}

\begin{figure*}
    \includegraphics[width=0.9\textwidth]{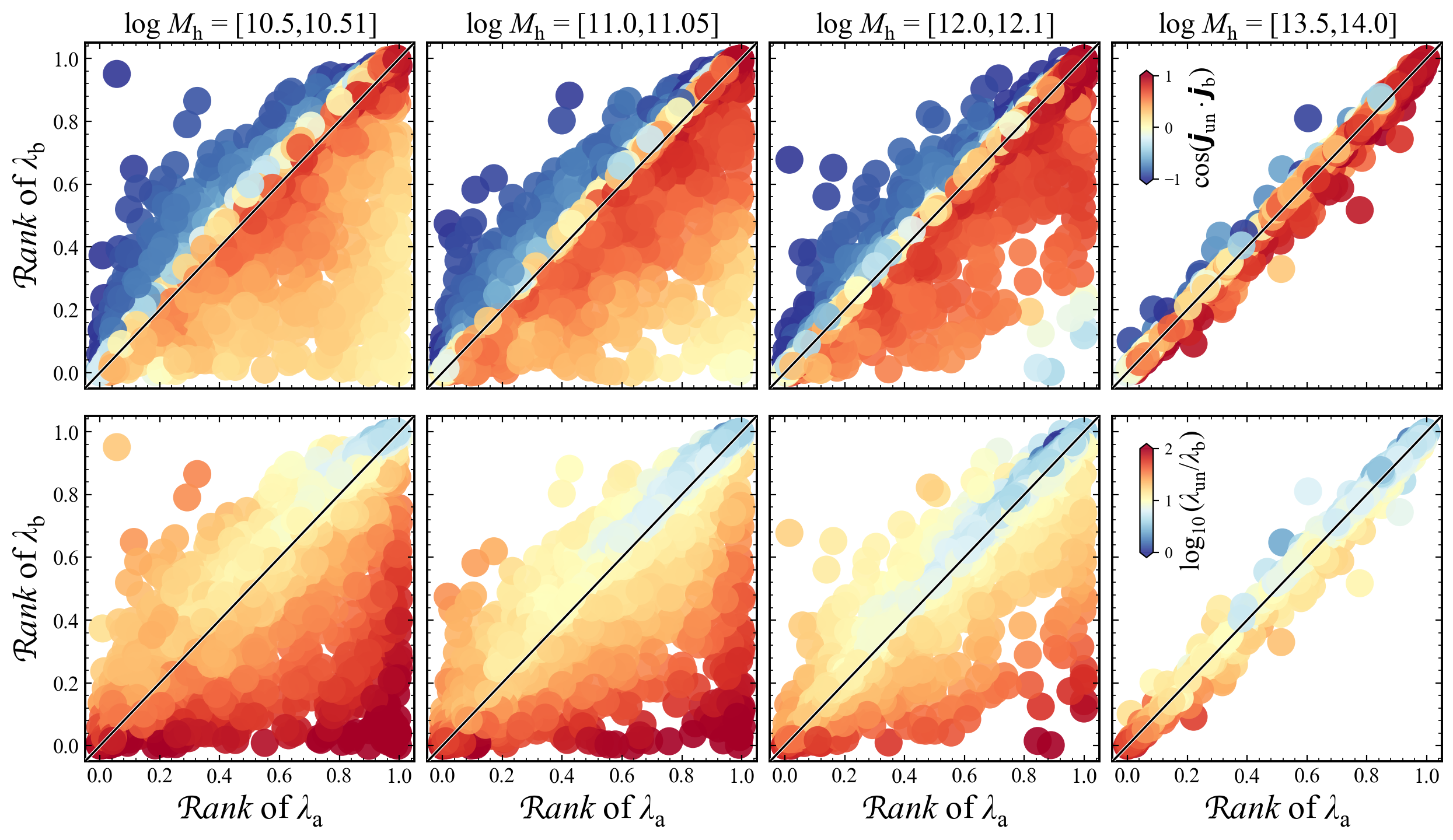}
    \caption{The ranking percentile of halo spin calculated by all particles ($\lambda_{\rm a}$) versus that calculated by bound particles ($\lambda_{\rm b}$) for different halo mass bins from left to the right panel. The color in upper panels indicates the misalignment between the angular momentum vector of the bound core, $\boldsymbol{j}_{\text{b}}$, and that of the unbound particles, $\boldsymbol{j}_{\text{un}}$. The color in lower panels indicates the difference between the spin parameter of the unbound particles ($\lambda_{\rm un}$) and that of the bound core ($\lambda_{\rm b}$). The color is smoothed by the LOESS method using 20\% nearest neighbors. Black line is the 1:1 relation. 
    }
    \label{fig:UDG_properties}
\end{figure*}

As outlined in \S\ref{sec:introduction}, recent high-resolution simulations have reported two conflicting trends 
in spin bias of halos at low masses, one showing that high-spin halos are more strongly clustered \citep[e.g.,][]{Montero-Dorta2020Manifestation,chenRelatingStructureDark2020,Wang2021EvaluatingOrigins}, while the other reveals an inversion 
below a characteristic mass of $\log M_{\rm h} [\rm M_{\odot}/h] \sim 11.5$ \citep[e.g.,][]{salcedoSpatialClusteringDark2018,sato-politoDependenceHaloBias2019,Johnson2019SecondarySpinBias, tucciPhysicalOriginsLowmass2021}. As noted by \citet{Montero-Dorta2020Manifestation}, this discrepancy could be caused by the different algorithms used for computing halo spins: 
studies in the former category use the Friends-of-Friends (FoF) algorithm while those in the latter use group finders  
such as {\tt SUBFIND} \citep{springelPopulatingClusterGalaxies2001} and {\tt ROCKSTAR} \citep{behrooziROCKSTARPHASESPACETEMPORAL2012}. 
The FoF algorithm typically uses all particles within the virial radius, whereas {\tt SUBFIND} and {\tt ROCKSTAR} use only 
bound particles. To resolve this discrepancy, we computed halo spin parameters using definitions with and without including 
unbound particles. We confirm that the spin bias inverts at a halo mass of $\sim 10^{11} h^{-1}{\rm M_{\odot}}$
when only bound particles are used. This crossover mass is approximately 0.5 dex lower than those found in 
previous studies. Our analyses comparing FoF and {\tt ROCKSTAR} halos reveal that this difference stems
largely from halo identification algorithms.  

Physically, unbound particles, which typically possess higher velocities than their bound counterparts, contribute significantly 
to a halo's total angular momentum. Furthermore, halos with a higher fraction of unbound particles, which usually
show larger disparity between $\lambda_{\rm a}$ and $\lambda_{\rm b}$, are predominantly located in denser regions, 
as shown in  Fig.~\ref{fig:Spin_compare}. This is consistent with the understanding that the unbound-particle fraction 
is influenced by the tidal field \citep{wangGalaxyClusteringProjected2011}: stronger tidal forces in denser environments promote tangential accretion and enhance particle acceleration \citep{Shi2015FlowPatternsaround}, thereby increasing both the unbound fraction 
and the overall halo spin. Consequently, spin measurements that include unbound particles ($\lambda_{\rm a}$) 
are elevated in such regions. 
This trend is particularly significant for low-mass halos which are more susceptible to environmental tidal influences.  

As suggested in \citealt{tucciPhysicalOriginsLowmass2021}, the splashback halos are a key population,  driving the low-mass spin bias. 
To assess the specific contribution of splashback halos to the spin bias discrepancy, we have repeated our analysis without removing them. We find that the spin bias for $\lambda_{\text{a}}$ remains more or less the same as that excluding splashback halos.  
For $\lambda_{\text{b}}$, however, the crossover mass decreases slightly, consistent with the trend reported by \citet{tucciPhysicalOriginsLowmass2021}. This indicates that, while splashback halos can induce the inversion of the 
spin bias measured with $\lambda_{\text{b}}$ at low masses, they make only a negligible contribution to the strong 
spin bias estimated using $\lambda_{\text{a}}$. 
Therefore, when splashback halos are removed to mimic the exclusion of red, high-S\'{e}rsic galaxies in the observational sample, the remaining spin bias is still strong enough to account for the clustering of UDGs. 
In other words, splashback halos are not the primary source for the unbound 
particles responsible for the enhanced spin and, consequently, for the strong clustering of UDGs.

To have a further insight into the relation between unbound particles and the density or tidal field, we examined both the magnitude and the orientation of the angular momentum of unbound particles relative to that of the bound particles in Fig.~\ref{fig:UDG_properties}.
There is a high degree of coherence between the angular momentum of bound particles and that of recently accreted unbound particles, due to the coherent infall along the filament \citep[e.g.][]{CodisConnectingCosmicWeb2012}.
Therefore, these halos exhibit a higher abundance in the lower-right corner in Fig.~\ref{fig:Spin_compare} and Fig.~\ref{fig:UDG_properties}.
However, for halos in very-dense environments (lower-right corner in Fig.~\ref{fig:Spin_compare}), we find the misalignment between the angular momentum of bound particles and that of unbound particles for these halos is nearly centered at zero. Instead, the unbound particles carry an extremely high value of the angular momentum. The magnitude of this momentum is sufficiently large that it effectively covers the original angular momentum of the system, regardless of the initial infalling direction. 
Furthermore, a rare population exists in the upper-left corner ($\lambda_{\text{b}} > \lambda_{\text{a}}$), characterized by a significant misalignment where the accreted material is counter-rotating relative to the bound core. These halos in high-density environments are similarly driven by the very high absolute velocity of the unbound particle flows.
The above findings suggest that the dense region (massive companion \citep{wangDistributionEjectedSubhaloes2009,sato-politoDependenceHaloBias2019} or the large-scale environment \citep
{lacernaNatureAssemblyBias2012,paranjapeHaloAssemblyBias2018, 
mansfieldThreeCausesLowmass2020,
Wang2024BeyondHaloMass,
MonteroDorta2024DependenceAssembly}, 
associated with strong tidal fields, can significantly enhance the halo spin by accelerating the unbound particle flows.



\subsection{The possible formation mechanisms of UDGs}

\begin{figure} \centering
    \includegraphics[width=0.4\textwidth]{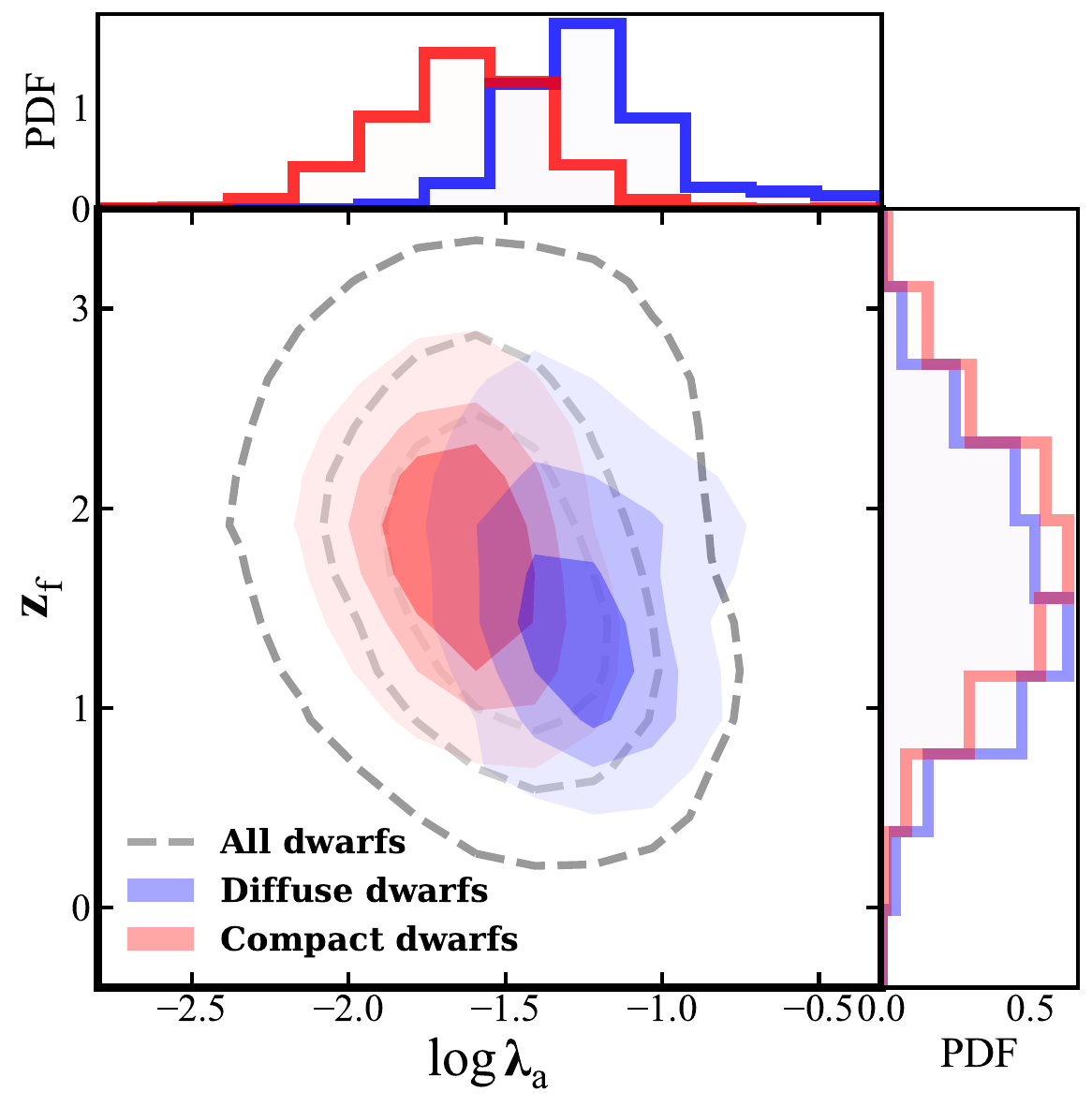}
    \caption{
        \textbf{The correlation between halo spin $\lambda_{\rm a}$ and halo formation time $z_{\rm f}$.}
        The blue and red contours are for the diffuse dwarf galaxies and compact dwarf galaxies subsample of the mock catalog with best fitting model with $\mathcal{R} = -0.54$, respectively.
        The colored contours cover 25\%, 50\% and 75\% number density of the mock catalog. The smaller panels show the one-dimensional distributions of the halo spin $\lambda_{\rm a}$ (top) and halo formation time $z_{\rm f}$ (right) for these two subsamples. 
        As a background, the black contours are for the whole dwarf galaxy sample, covering 50\%, 75\% and 90\% number density, respectively.
    }
    \label{fig:spin_zf}
\end{figure}

\citetalias{Zhang2025UnexpectedClustering} studies the spin bias using a spin parameter computed using 
{\tt SUBFIND} (and thus based on bound particles only), so their finding that this spin bias cannot reproduce 
the strong UDG clustering is consistent with our results. The authors then turned to halo assembly bias in terms of 
formation time, finding that their observations could only be reproduced if diffuse dwarf galaxies reside in older 
halos, namely if $\Sigma_\ast$ is anti-correlated with halo formation redshift $z_{\rm f}$ (defined as the redshift 
when the main progenitor of a halo first reaches half its present-day mass; \citealt{gaoAgeDependenceHalo2005}). 
Since this anti-correlation is not seen in current hydrodynamical simulations and semi-analytic models, \citetalias{Zhang2025UnexpectedClustering} invoked self-interacting dark matter (SIDM) as a potential 
explanation.

To examine the role of halo formation time ($z_{\rm f}$) in our model, Fig.~\ref{fig:spin_zf} shows the relationship between $\lambda_{\rm a}$ and $z_{\rm f}$ for the most compact and the most diffuse dwarf galaxies in our best-fit model. The side panels show the one-dimensional distributions of $\lambda_{\rm a}$ (top) and $z_{\rm f}$ (right). The diffuse and compact galaxies show significantly different distributions  of $\lambda_{\text{a}}$, as implied by our model construction, but have substantial overlap in the $z_{\rm f}$ distribution. 
The diffuse galaxy subsample has a slightly smaller median $z_{\rm f}$ (i.e., younger halos) than the compact subsample. 
Therefore, our model demonstrates that the unexpected clustering of dwarf galaxies can be explained by the spin bias,
through the inclusion of unbound particles, without requiring diffuse galaxies to reside in older halos or invoking 
non-standard dark matter.


\begin{figure*} \centering
    \includegraphics[width=0.97\textwidth]{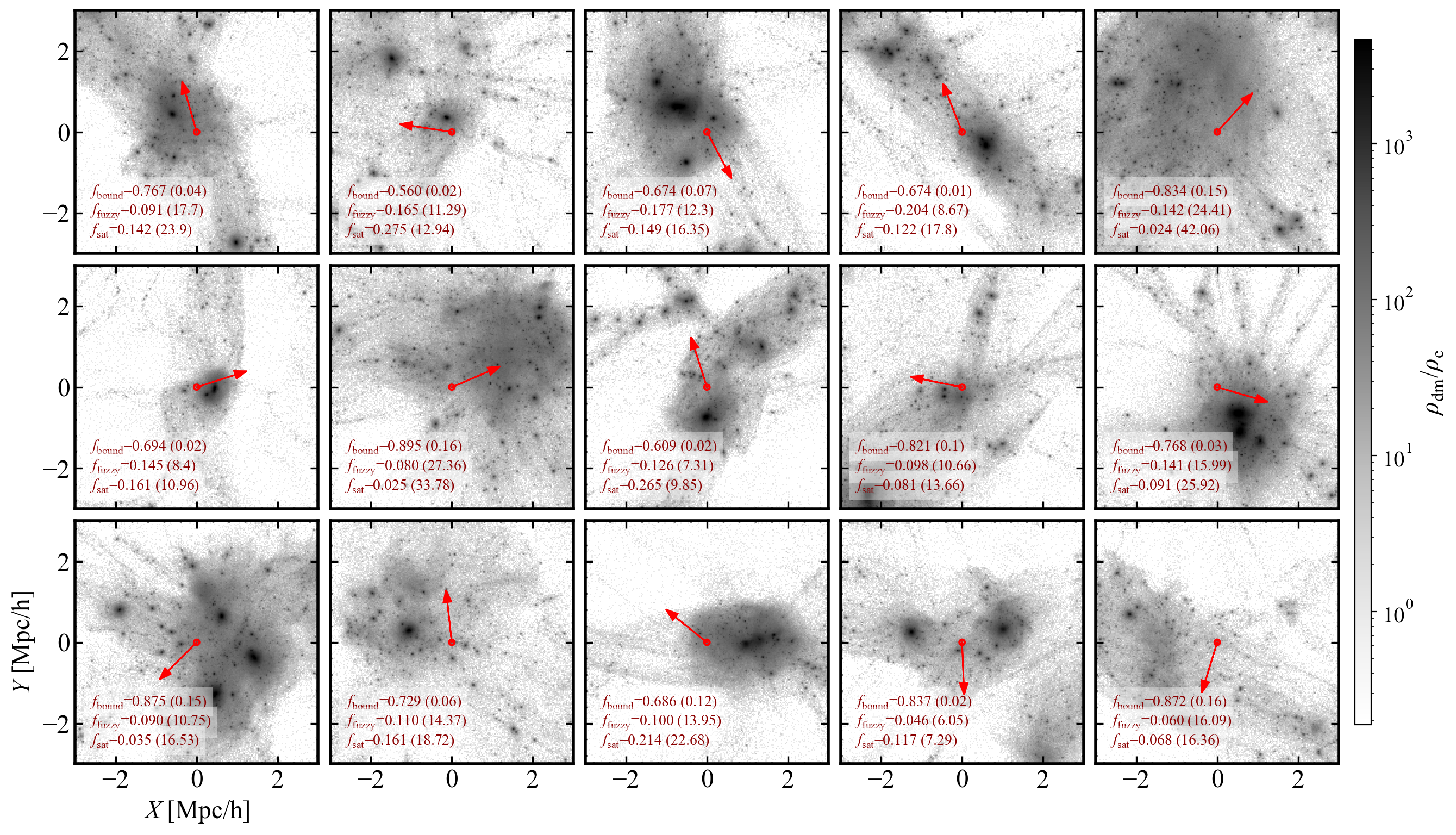}
    \caption{
        \textbf{Environment of 15 example dwarf-host halos with high spin.}        
        For each panel, it shows the projected dark matter density ($\rho_{\rm dm}$) centered on a high-spin halo, averaged over $\pm 0.5 \rm \ Mpc/h$ along the $z$-axis, normalized by the critical density of the universe. The red dot marks the halo's virial radius ($R_{\rm vir}$), and the red arrow shows the direction of the velocity of the halo. The lower-left box displays key metrics: $f_{\rm bound}$, the mass fraction of bound particles in all particles within $R_{\rm vir}$; $f_{\rm fuzzy}$, the fraction of fuzzy unbound particles among the all particles; and $f_{\rm sat}$, the fraction of unbound particles belonging to satellites among the all particles. The parenthetical values represent the mean velocity of each population (relative to the high-spin halo) normalized by the halo's virial velocity.
    }
    \label{fig:unbound_source}
\end{figure*}

To gain further insight into how unbound particles contribute to the formation of diffuse dwarf galaxies, we
investigate the origin of the unbound particles by selecting 15 halos whose spin parameters ($\lambda_{\rm a} \geq 0.01$) are 
among the highest 0.1\% of all halos with $M_h \in 10^{10.5} \rm M_{\odot}/h$. 
We show the density field surrounding each of these halos in one of the panels of Fig.~\ref{fig:unbound_source}. 
All these halos are close to massive structures, being either 
a massive halo or a dense filament connecting massive halos.
This aligns with the finding by \citetalias{Zhang2025UnexpectedClustering} (see their Fig. 2f) 
that diffuse dwarfs (mostly UDGs) tend to be located around massive knots 
and filaments in the cosmic web, thus providing another support to 
our expectation that these halos can host diffuse dwarfs.
However, only a few of these halos are moving towards the nearby massive halos,
as seen from the red arrows indicating their velocity directions.
This suggests that the high-spin halos are not falling towards the 
massive halos, but rather passing by the over-densities (filaments or 
the outskirts of knots) at high speeds. A passing-by halo is thus 
expected to face a headwind of dark matter, with a fraction of the wind particles
mixed into (but not bound to) the halo and contributing to the population of 
unbound particles. If the dark-matter wind also has a large impact parameter
with respect to the halo center, which is found to be the case
as unbound particles tend to reside around the outskirts  of halos,
it may inject a significant amount of angular momentum to the halo if it can be 
harnessed.   

To further illustrate this point, we classify particles within its $R_{\rm h}$ 
in each halo into three subpopulations: (i) those bound to the halo; (ii) those not bound to the halo 
but bound to satellite subhalos; (iii) those not bound to the halo and its subhalos.
We denote these subpopulations as bound, satellite and fuzzy particles, 
respectively, and we list their mass fractions, as well as their 
mean velocities relative to the halo, in each panel of Fig.~\ref{fig:unbound_source}.
The fraction of unbound particles (satellite + fuzzy) is around 20\%,
and in many cases it exceeds 30\%. 
In contrast, our examination of the entire halo population 
with $M_{\rm h}\sim 10^{10.5}\, \rm M_{\odot}$ showed that their unbound
fraction is about 10\%. Satellite and fuzzy subpopulations contribute comparably 
to the unbound particles in high-spin halos, and both have velocities of 
$\gtrsim 10$ times the $V_{\rm h}$ of the halo.
Such high velocities thus confirm the passing-by scenario proposed above,
and, together with the high fractions of unbound particles in high-spin halos, 
explain the significant increase of the halo spin.

The high-speed, off-center dark-matter wind (either fuzzy particles or those stripped from satellites)
faced by a passing-by halo is expected to have important implications for the gas content in the halo.
The gas associated with the dark-matter wind can interact and mix with 
the preexisting gas in the halo through ram pressure, producing shocks
and transferring angular momentum into the halo gas through ram pressure torquing. 
The typical velocity of high-spin halos  ($\gtrsim 10 V_{\rm h}$) 
relative to the wind implies that the turbulence introduced by the shock is initially 
supersonic, with a Mach number of $\mathcal{M}_{\rm s} \gtrsim 10$, 
and is thus compressive in nature.
The high density of cosmic filaments, with a typical value of about $10$ times the cosmic 
mean \citep{cautunEvolutionCosmicWeb2014},
together with the high speed of $\gtrsim 10V_{\rm h}$,
implies that the amount of gas swept by the halo within a dynamical
timescale of the halo is comparable to total amount of 
baryon of the halo itself, and that the supersonic turbulence may 
thus be sustained to continuously compress the halo gas.
The supersonic shock initially introduces a density enhancement in the halo gas. This compression, even at the adiabatic limit (a factor of 4 for $\gamma=5/3$), can significantly shorten the local cooling timescale. 
Given the already high density of the cosmic filaments, this shock-induced perturbation is expected to trigger runaway radiative cooling. As the shock-heated gas radiates its thermal energy, the shock transitions toward an isothermal state, where the post-shock density can be further elevated by a factor as large as $\mathcal{M}_s^2 \gtrsim 100$, effectively sustaining the cold gas inflow. 
The cooled gas can thus flow into the central galaxy, carrying in a large amount of 
angular momentum that originated from the wind and has been mixed into 
the halo gas during the shock.
The post-passing-by process may thus lead to the formation of an extended 
disk population, producing a diffuse galaxy.

The passing-by scenario proposed here, of course, needs to be revisited and 
verified by hydrodynamical simulations that implement the physical processes
required to resolve the aforementioned processes.
Some evidence has been found in recent hydrodynamical simulations. For example, 
\citet{liShockinducedStrippingSatellite2023} found that the ISM in some galaxies 
in the TNG simulation facing ram pressure is not significantly affected and, 
in extreme cases, is compressed.
\citet{duFormationCompactElliptical2019} found that the ram pressure in dense 
environment can lead to a divergent effect composed of both stripping 
and confinement. 
The key requirement for our scenario to be simulated is that the mixing between 
gas in the passing-by halo and gas associated with the dark-matter wind
should be resolved properly and that the subsequent cooling, inflowing, and star formation 
should also be captured correctly.

\subsection{Implications for galaxy formation models and observations of halo secondary bias}

The strong correlation between dwarf galaxy size and halo spin (calculated using all particles) has significant 
implications for both semi-analytical models (SAMs) and hydrodynamic simulations. SAMs typically assume that
disk size is proportional to the product of halo spin and virial radius 
\citep[e.g.][]{somervilleSemianalyticModelCoevolution2008, bensonALACTICUSSemianalyticModel2012}. 
Our results suggest that to best recover the properties of dwarf galaxies, these models should also consider a spin 
definition that includes unbound particles. While this does not affect the overall size distribution, it can substantially change 
the spin-ranking of individual halos.

In hydrodynamic simulations, although current simulations are powerful enough to reproduce basic properties 
(e.g., mass and size distributions) of dwarf galaxies \citep[e.g.][]{Sales2022BaryonicSolutionsChallenges}, 
the formation scenarios considered are often diverse. 
These include the halo spin scenario  \citep[IllustrisTNG and Auriga;][]{liaoUltradiffuseGalaxiesAuriga2019,benavidesOriginEvolutionUltradiffuse2023}, 
the galactic fountain model \citep{Zheng2025UltradiffuseGalaxies}, 
and models based on major mergers \citep{wrightFormationIsolatedUltradiffuse2021} and stellar feedback-driven outflows \citep{dicintioNIHAOXIFormation2017,chanOriginUltraDiffuse2018,martinFormationEvolutionLowsurfacebrightness2019,Freundlich2020ModelCore}.
The TNG50 and Auriga simulations, which support the halo spin scenario, show a strong correlation between halo spin and gas spin \citep{Yanggalaxysizehalospin2023,liangConnectionGalaxyMorphology2024}, while other simulations show weak or no correlation \citep{Desmond2017GalaxyhaloEAGLE, jiangDarkmatterHaloSpin2019,Yanggalaxysizehalospin2023}.
All these indicate that the transfer of angular momentum from the dark matter halo to the gas is different across 
simulations, presumably because such transfer is sensitive to the treatment of feedback-driven outflows and the recycled gas 
fraction \citep{Ublerfeedback2014,AgertzFeedbackongalaxy2016,zjupaAngularMomentumProperties2017,Yang2024ApostleaurigaEffects}
and the emergent outcomes of the different treatments for sub-grid processes are different.
Regarding this point, our results may provide some insight into the relationship between halo spin and gas 
spin for dwarf galaxies and help constrain the sub-grid implementations in simulations.

Furthermore, the connection between galaxy size and halo spin offers a unique opportunity to measure halo spin bias observationally. As UDGs in our model are associated with high-spin halos, their "unexpected" clustering—as reported in \citetalias{Zhang2025UnexpectedClustering}—serves as a potential proxy for the secondary bias of halos. Dwarf galaxies are ideal candidates for such a study because they are less susceptible to halo mass estimation errors and exhibit more pronounced secondary bias effects than massive systems. By using galaxy size as a reliable proxy for halo spin ($\lambda_{\rm a}$), we can bridge the gap between theoretical predictions of assembly bias and large-scale structure observations.

We note that our model is purely empirical, as it relies solely on the halo spin at $z = 0$, and detailed models are required to fully understand 
the formation of dwarf galaxies. On the observational side, the clustering properties of UDGs warrant further investigation. 
For instance, dwarf galaxies identified in SDSS are primarily located within local voids, which may introduce significant cosmic 
variance into their clustering signals. This issue requires larger and deeper surveys, such as DESI and LSST.
Furthermore, SDSS predominantly probes the central regions of these galaxies, and the low-surface density observed in the core 
may not necessarily imply that the entire system qualifies as an ultra-diffuse galaxy. Additionally, the sample of late-type, blue, 
isolated dwarf galaxies may still be subject to potential systematic biases.
Further investigations on these fronts are therefore essential to fully understand the implications of the results found here.

\section{Summary}
\label{sec:conclusion}
We have used the TNG300-1-Dark simulation to investigate the spin bias of low-mass halos and applied our findings to explain the strong clustering of ultra-diffuse galaxies (UDGs) recently reported by \citetalias{Zhang2025UnexpectedClustering} from SDSS. We evaluated two definitions of halo spin: one using only gravitationally bound particles within the virial radius ($\lambda_{\rm b}$), and another that includes unbound particles ($\lambda_{\rm a}$).

We find that while the overall spin distribution depends only weakly on the definition, the inferred spin bias is highly sensitive to it, particularly for low-mass halos. The spin bias measured with $\lambda_{\rm a}$ shows that higher-spin halos are more strongly clustered across all masses, whereas the bias measured with $\lambda_{\rm b}$ inverts below $M_{\rm h}\sim 10^{11} \rm M_{\odot}/h$, with lower-spin halos becoming more clustered. This discrepancy arises because a subset of low-mass halos exhibits large $\lambda_{\rm a}$ but simultaneously small $\lambda_{\rm b}$. These halos are predominantly found in higher-density environments, which drives the significant difference in spin bias at low masses.

We developed an empirical model, generating SDSS-like mock catalogs to link dwarf galaxy surface mass density ($\Sigma_\ast$) with the halo spin parameter $\lambda_{\rm a}$. Assuming a monotonic $\Sigma_\ast$-$\lambda_{\rm a}$ relation with scatter, our model—constrained by the observed dwarf galaxy clustering from \citetalias{Zhang2025UnexpectedClustering}—reveals a significant anti-correlation. This indicates that more diffuse dwarf galaxies indeed form in higher-spin halos. Consequently, the strong clustering of dwarf galaxies can be naturally explained by spin bias within the standard $\Lambda$CDM framework, without resorting to non-standard dark matter models such as the self-interacting dark matter proposed by \citetalias{Zhang2025UnexpectedClustering}.



\section*{Data Availability}

TNG300-1-Dark is available at 
\url{https://www.tng-project.org}. {\software Dwarf-bias} is available at 
\url{https://github.com/ChenYangyao/dwarf_assembly_bias}.

\section*{Acknowledgements}

We are grateful to the anonymous referee whose comments have helped us to improve this article.
QM acknowledges Simon White, Chenyang Ji, and Leyao Wei for discussion.
This work is supported by the National Key R\&D Program of China (grant NO. 2022YFA1602902), the National Natural Science Foundation of China (grant Nos. 12433003, 11821303, 11973030), and the China Manned Space Program with grant no. CMS-CSST-2025-A10.

\bibliographystyle{mnras}
\bibliography{references} 

\appendix

\bsp	
\label{lastpage}
\end{document}